\documentclass[aps,prl,twocolumn,showpacs,superscriptaddress,floatfix]{revtex4-1}
\usepackage[utf8]{inputenc}
\usepackage{amsmath}
\usepackage{amsfonts}
\usepackage{graphicx}
\usepackage{braket}
\usepackage{dsfont}
\usepackage{color}
\usepackage{sidecap}
\usepackage[dvipsnames]{xcolor}

\def \beq {\begin{eqnarray}}
\def \eeq {\end{eqnarray}}
\def \Schrodinger {{Schr\"{o}dinger }}

\begin{document}
\title{Time propagation and spectroscopy of Fermionic systems using a stochastic technique} 
\author{Kai Guther}
\email{k.guther@fkf.mpg.de}
\affiliation{Max Planck Institute for Solid State Research, Heisenbergstra{\ss}e 1, 70569 Stuttgart, Germany}
\author{Werner Dobrautz}
\email{w.dobrautz@fkf.mpg.de}
\affiliation{Max Planck Institute for Solid State Research, Heisenbergstra{\ss}e 1, 70569 Stuttgart, Germany}
\author{Olle Gunnarsson}
\email{o.gunnarsson@fkf.mpg.de}
\affiliation{Max Planck Institute for Solid State Research, Heisenbergstra{\ss}e 1, 70569 Stuttgart, Germany}
\author{Ali Alavi}
\email{a.alavi@fkf.mpg.de}
\affiliation{Max Planck Institute for Solid State Research, Heisenbergstra{\ss}e 1, 70569 Stuttgart, Germany}
\affiliation{University Chemical Laboratory, Lensfield Road, Cambridge, CB2 1EW, U.K.}

\begin{abstract}
We present a stochastic method for solving the time-dependent Schr\"odinger equation,
generalizing a ground state full configuration interaction Quantum Monte Carlo method. 
By performing the time-integration in the complex plane close to the real time axis, the numerical
effort is kept manageable and the analytic continuation to real frequencies is efficient.
This allows us to perform {\it ab initio} calculation of electron spectra for
strongly correlated systems. The method can be used as cluster solver for embedding schemes.
\end{abstract}
\maketitle
{\it Introduction.}
The time evolution of a closed interacting electronic system, having been prepared in a well-defined but entangled non-stationary 
state, is of considerable interest to a broad range of fields. This includes many types of electronic 
spectroscopy such as photoemission (PE) and inverse photoemission (IPE) \cite{HL1969,T1970,
  HF1984} , core-level \cite{PSBM1976, MS1983} and optical spectroscopies, as well as the field of 
non-equilibrium dynamics \cite{PSAM2011}, including dynamics in driven, time-dependent, external fields.
In solid-state physics, such electronic spectroscopies play a leading role in providing information on the 
electronic structure of the material. In weakly-correlated materials, the
GW-approximation provides a viable theoretical tool for calculating excitation
energies \cite{HL1969, ORR2002}. In strongly-correlated materials, however,  
theoretical studies are often limited to model systems such as the Hubbard \cite{Hubbard} or 
Anderson \cite{Anderson} models. 
Efficient methods have been developed for studying such models.\cite{dca,continuoustime}
However, it is not clear how these methods can be generalized to {\it ab
  initio} calculations. Here we show how this can be achieved using a time evolution method stochastically applied to {\em ab initio} Hamiltonians.  

Time evolution of quantum systems is a notoriously difficult problem owing to the existence of a severe dynamical 
sign problem. For electronic systems there is another difficult sign-problem due to its fermionic nature. Fundamentally, we are required to integrate 
the time-dependent \Schrodinger equation for a many-electron system for long times. 
Methods based on deterministic wavefunction propagation, such as the Crank-Nicolson method \cite{CrankNicolson1947}, or 
Lanczos recursion \cite{Parlett,ParkLight1986}, suffer from severe memory requirements. 
Quantum Monte Carlo methods (especially quantum lattice methods)   
typically work in imaginary frequency space \cite{dca, continuoustime}, followed by analytic continuation to to real frequencies. The analytic continuation 
is numerically highly ill-conditioned, and maximum entropy (MaxEnt) methods \cite{maxent,Jarrell} are usually employed.
Although spectral features close to the Fermi energy can be obtained rather accurately, features 
further away, e.g., satellites, are smeared out (see appendix \cite{supplement}).
Such satellites, however, can contain a wealth of information about the dynamics of the system.
In ab initio models these problems are further exacerbated by the large range of energies spanned by the basis set 
(over numerous Hartrees) and the huge Hilbert spaces owing to the large number
of virtual orbitals.    

In this letter we present an approach to this problem. We present a real-time generalization of an algorithm for  
calculating  fermionic ground states using imaginary-time propagation. This involves the introduction of a second-order 
time propagator, which is implemented in a stochastic manner.
This approach yields accurate time-correlation functions, but 
the computational cost increases exponentially, as the undamped time-evolving wave functions explores the available (exponentially large) 
Hilbert space. To ameliorate this problem, we introduce an adaptive  
variable-phase time-step into the propagator, which leads to a propagation in
the complex plane close to the real time axis. This results in a slow damping, which keeps the computational cost essentially fixed (similar
to a ground state calculation). Nevertheless, this gives phase information about the wave function and yields oscillatory time-correlation functions. 
We have developed a MaxEnt scheme, which performs analytic continuation from an arbitrary path in complex time space
to real frequencies. This provides spectral functions over a broad energy range. 
We apply the method to benchmark systems for which numerically exact results are available, and show that these 
are reproduced to high accuracy at a fraction of the cost. Then we apply the algorithm to ab initio (atomic and molecular)
systems, where comparison is made with experiment. 

In {\it ab initio} calculations for solids, this method could be used as a
cluster solver in embedding schemes like dynamical cluster approximation\cite{dca}.

{\it Real-time evolution.}
\label{sec:rneci}
Given a Hamiltonian $\hat{H}$ and an initial wave function $\Ket{\Psi(0)}$, we wish to solve                          
the time-dependent \Schrodinger equation: 
\begin{equation}\label{eq:1}
\mathrm{i}{\partial \over \partial t}|\Psi(t)\rangle  =  \hat{H}|\Psi(t)\rangle
\end{equation}
$|\Psi(t)\rangle$ gives information about various spectroscopic properties. We can see this by considering the
inverse photoemission spectrum $A_{ii}(\omega)$ 
\begin{equation}\label{eq:2}
A_{ii}(\omega)=\sum_n|\langle\Psi_n^{N+1}|c_{i\sigma}^{\dagger}|\Psi_0^N\rangle|^2\delta(\omega-E_n^{N+1}+E_0^{N}+\mu),
\end{equation}
where $c_{i\sigma}^{\dagger}$ adds an electron with spin $\sigma$ to  orbital $i$ in the ground state $|\Psi_0^N\rangle$ with $N$ electrons. Here $|\Psi_n^{N+1}\rangle$
is the $n$th excited state of the $(N+1)$-electron system.  $E_0^{N}$ and $E_n^{N+1}$ are the corresponding
energies and $\mu$ is the chemical potential. The formal solution of Eq.~(\ref{eq:1}) is $|\Psi(t)\rangle =
{\rm exp}(-i\hat H t)|\Psi(0)\rangle\equiv \hat U(t)|\Psi(0)\rangle$.
The spectrum is then given by
\begin{equation}\label{eq:3}
A_{ii}(\omega)={1\over \pi}{\rm Im} \left[-\mathrm{i}\int_0^{\infty}dt e^{i[\omega+i0^{+}+E_0(N)+\mu]t}\langle \Psi(0)|\Psi(t)\rangle\right],
\end{equation}
where we have used the initial condition
$\Ket{\Psi(0)}=c_{i\sigma}^{\dagger}|\Psi_0^N\rangle$ and $0^{+}$ is a positive
infinitesimal quantity and the calculated object is the Green's function $\Braket{\Psi(0)|\Psi(t)}$.
In a similar way the photoemission spectrum can be calculated. These formulas
are discussed in detail in the appendix \cite{supplement}.

{\it Methods.} To compute $|\Psi(t) \rangle$ accurately for long propagation times,   
we have adapted the Full Configuration Interaction Quantum Monte Carlo (FCIQMC) method \cite{BAT2009,BA2010,CBA2010,BGKA2012}.
This method was originally designed to stochastically project the wave function,
expressed in a full Slater determinant basis $\{\ket{D_i}\}$,  towards the ground state. 
The ground state algorithm uses a stochastic representation of the full CI wave function $\Psi=\sum_i C_i \ket{D_i}$ using 
signed walkers, $C_i$, together with the repeated stochastic application of a short-time propagator $\hat{P}(\Delta \tau) 
= \mathds{1} - \Delta \tau \hat{H}$ to the population of walkers, followed by walker annihilation at the end of each iteration. 
More details are given in the appendix \cite{supplement}.  

\begin{SCfigure*}
\centering
\caption{(a) Time evolution of Re $\langle \Psi(0)|\Psi(t)\rangle $ and
  contour in complex time and (b) 
corresponding photoemission spectra (for $\mu=0$) for 
 the time-evolution using 70000, $1.6\times 10^6$  and $1.7\times10^7$  walkers for the 18-site Hubbard model at
  $U/t=2$, $k=(0,0)$ and half-filling. All calculations start from the same initial state with 350000
walkers, and three different time contours were used leading to 70000,
$1.6\times 10^6$  and $1.7\times10^7$ walkers for
longer times. The utilized time-step is $10^{-3}$. Both the Lanczos and FCIQMC spectra were convoluted with a Lorentzian of
full width at half maximum (FWHM) of $0.02$ to simplify visual comparison of  the FCIQMC spectrum to the discrete
eigenvalues obtained in the Lanczos method. The integrated weights of the
peaks of the FCIQMC spectra are indicated and agree well with the weights of
the discrete Lanczos spectrum, which are given in the first graph of b). The
bracketed numbers indicate the weights of not fully resolved peaks.
(c) Photoemission and inverse photoemission spectra for a 24-site
cluster with lattice vectors (3,3) and (-5,3) with 22 electrons at $U/t=4$ for
$k=(0,0)$ obtained using $\sim 1.5\times 10^8$ and $\sim 3\times 10^7$ walkers respectively. The inverse photoemission part carries very low weight and is also
shown in the inset. For comparison, the same spectrum
computed by means of the Hirsch-Fye\cite{HF} auxiliary-field quantum Monte Carlo (AFQMC) is
displayed.
}
\includegraphics[width=0.9\columnwidth]{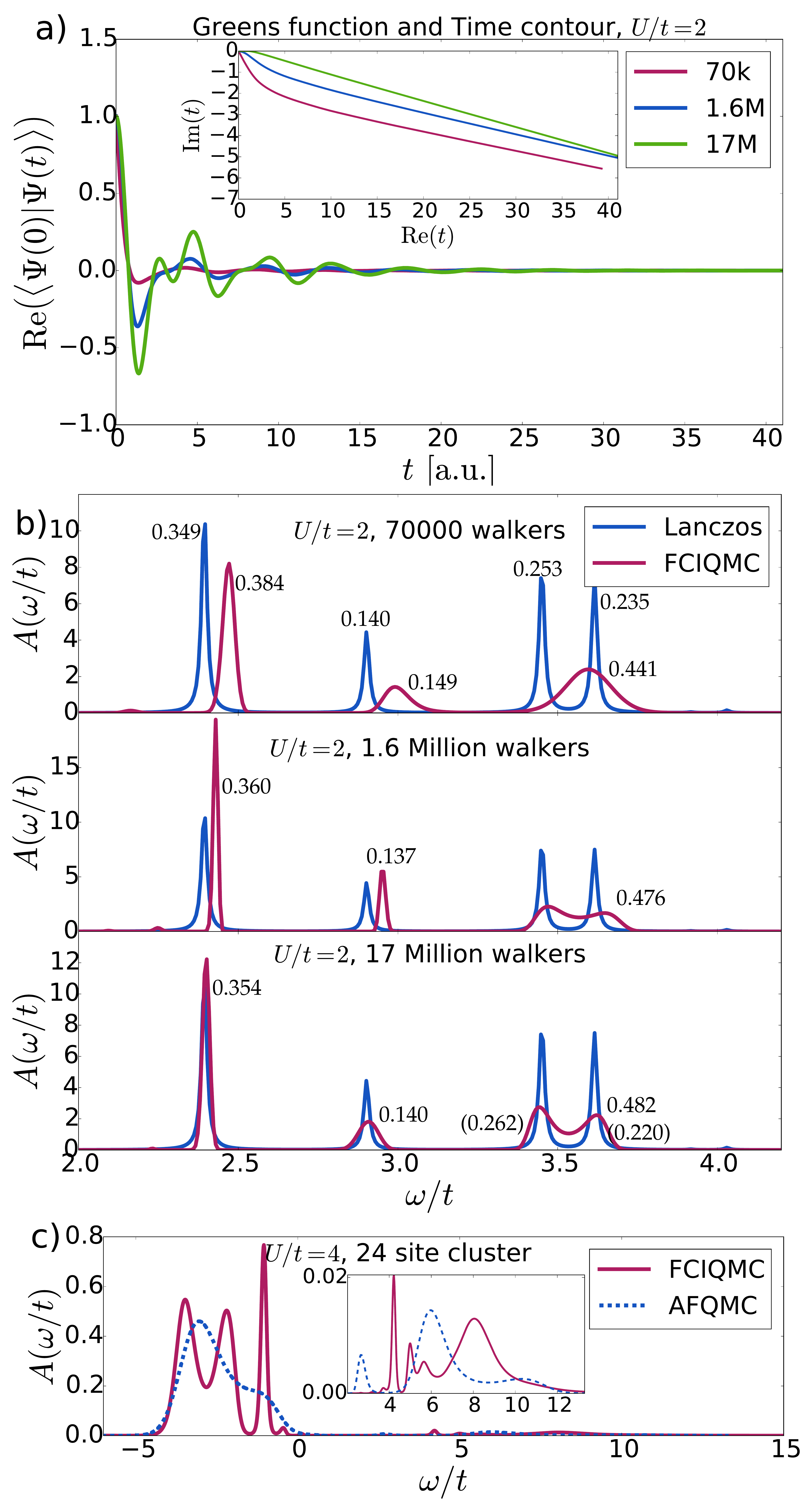}
\label{fig:panelHubWn}
\end{SCfigure*}

Generalizing to the time dependent problem, the wave function $\Psi(t)=\sum_i C_i(t) \ket{D_i}$ is represented by a collection 
of complex walkers, the time evolution of which is realized through the  
successive application of a {\em second-order} propagator: 
\begin{equation}
\hat{U}_2(\Delta t) = \mathds{1} - \mathrm{i}\Delta t\hat{H} -
\frac{1}{2}\left(\Delta t\right)^2 \hat{H}^2  \,.
\end{equation}
where $\Delta t$ is a small time-step. Thus $\Psi(t+\Delta t) = \hat{U}_2(\Delta t) \Psi(t)$. 
This approach preserves the norm of the wave function to 
order ${\cal O}(\Delta t^4)$ per step and ${\cal O}(\Delta t^3)$ in total,
which is found to be sufficient to allow for stable propagation for a long
time, without significant norm-conservation errors. In contrast, propagation 
using  a first-order propagator only leads to norm-conservation of order 
$\Delta t$, which leads to a severe violation of unitarity over relevant 
time-scales. The time evolution is implemented using a second-order Runge-Kutta
algorithm. Numerical examples are provided in the appendix
\cite{supplement}. 

Although this method remains unitary to a good approximation, 
stochastic errors lead to a growth of the norm over time
(see appendix \cite{supplement}), which becomes 
unmanageable for large Hilbert spaces.
We therefore allow the time step $\Delta t$ to acquire a phase $\alpha$ 
\begin{equation}
\Delta t \mapsto \,\mathrm{e}^{-\mathrm{i}\alpha}\Delta t\,,
\end{equation}
thereby introducing a damping in the propagator. The phase is varied
dynamically to keep the number of walkers approximately constant. 
A small number of walkers requires a large $\alpha$, and increasing the 
number of walkers reduces $\alpha$. The pure real-time propagation  
($\alpha=0$) is achieved in the large walker limit. Since $\alpha\ne 0$ 
results in complex-time Green's functions,  we have generalized the 
(imaginary time) MaxEnt method \cite{maxent,Jarrell} to compute $A(\omega)$ 
(see appendix \cite{supplement}). The analytic continuation is 
more accurate for small $\alpha$, and robustness of the calculated spectra can be 
checked by comparing results for different numbers of walkers. To obtain 
the statistics needed for the MaxEnt method, we run several independent calculations.

Compared with the finite temperature Matsubara (imaginary time) formalism, 
this leads to three advantages. i) The MaxEnt method gives a more
detailed spectrum, since the time path is rather close to the real axis,
rather than along the imaginary axis. ii) In each spectral calculation we shift $\mu$ so that the peak closest to $\mu$ is located at $\mu$. Since MaxEnt  is most accurate close to
$\mu$, this improves the accuracy. iii) For a given ${\bf k}$, the weight of the 
PE and IPE spectra may be very different. By performing the PE and IPE
calculations separately, we obtain a comparable relative standard deviation 
in both cases, in contrast to the Matsubara formulation. These aspects are 
discussed in the appendix \cite{supplement} and illustrated in
Fig.~\ref{fig:panelHubWn}c below.

{\it Application to the Hubbard model.}
As a first example, we consider the fermionic Hubbard model \cite{Hubbard}. It is defined by the Hamiltonian 
$               
  H = -t\sum_{\langle i,j\rangle \sigma}         c^{\dagger}_{i\sigma} c^{}_{j \sigma} + U \sum_i
  n_{i\uparrow} n_{i\downarrow} \,.
$
We consider a two-dimensional square-lattice with periodic boundary conditions.  

We apply the method to an 18-site cluster (18A in Betts' notation\cite{Bett}) at half-filling, 
which is among the largest Hubbard systems whose Green's function can be calculated 
numerically exactly using Lanczos recursion \cite{Parlett,Gunnarsson1986} 
(with a Hilbert space  consisting of $\sim 2.4\times 10^9$ determinants).
To compute the Green's function, we first converge the ground state using 
imaginary-time FCIQMC, and then perform a complex time calculation with
a ${\bf k}=(0,0)$ electron removed from the ground state. A plane waves basis set is  used
here.

\begin{figure}[t!]
\centering
\includegraphics[width=0.9\columnwidth]{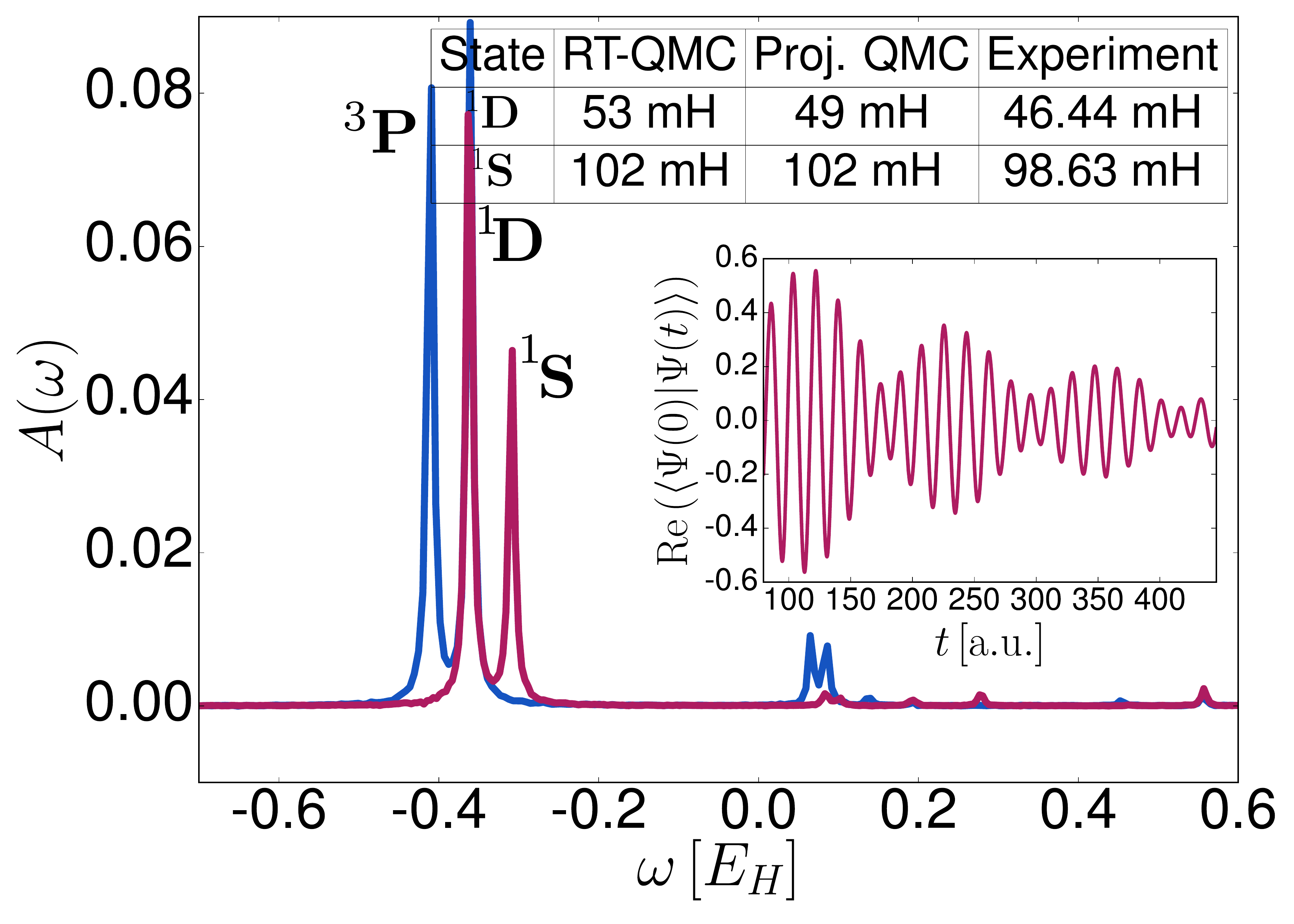}
\caption{Atomic multiplet of the carbon atom, obtained from two distinct initial states created by
  adding a 2p electron to the cation ground state. One of the states is
  prepared as a singlet (red), the second state (blue) is a mixture of singlet and
  triplet but with $L_z\neq 0$. The time-evolution is carried out
  for 1600 a.u. of time and the zero of the frequency axis corresponds
to $-37.3706\,\mathrm{H}$ which is the ground state energy of the cation
computed using the projective FCIQMC algorithm. The VTZ basis is used in this example. 
The experimental values are
according to \cite{KW1966}. The frequency resolution is $3.9\,\mathrm{mH}$. The inset shows a portion of the computed Green's function in real time. 
}
\label{fig:panelCarbonStats}
\end{figure}

\begin{figure}[t!]
\centering
\includegraphics[width=\columnwidth]{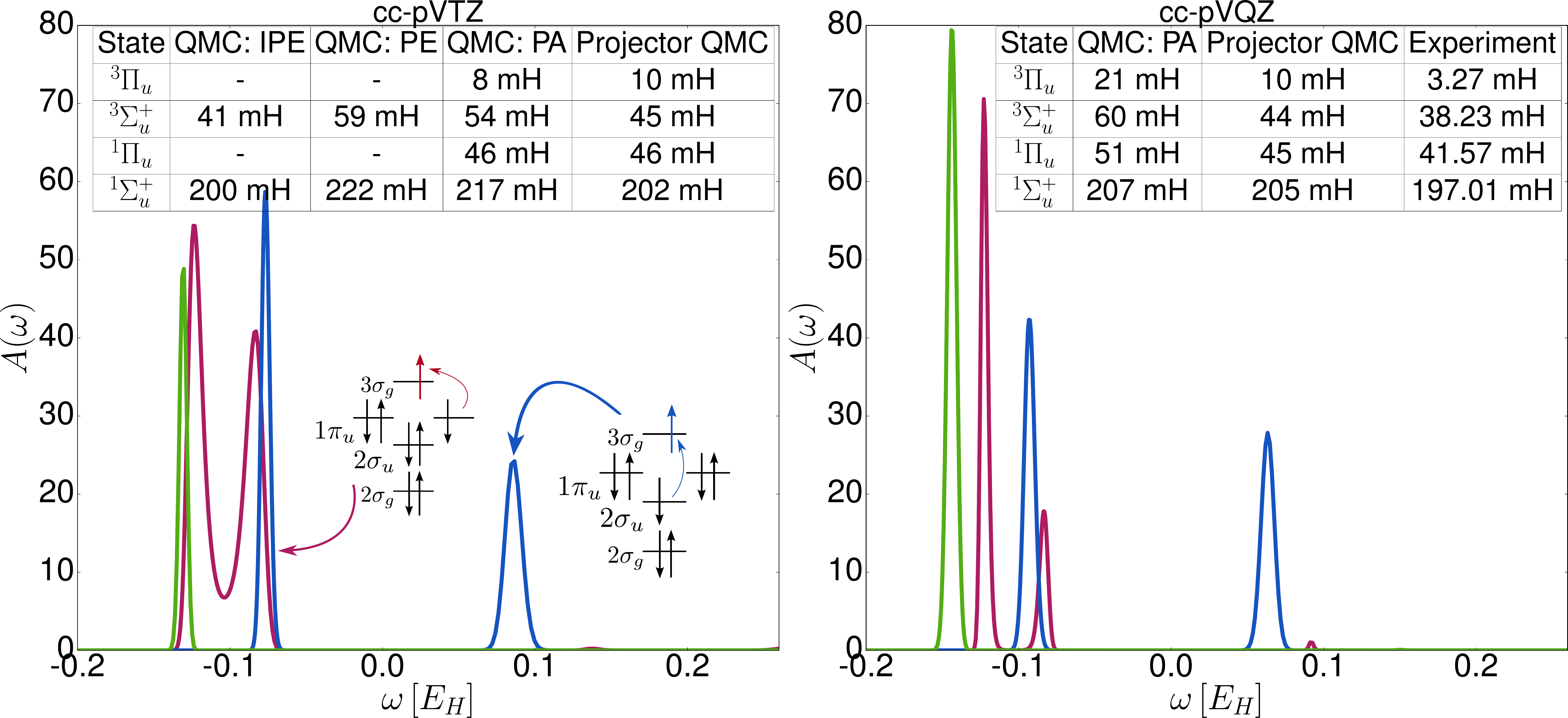}
\caption{Photoabsorption spectra for the carbon dimer for a single excitation
  from the $2\sigma_u$ to the $3\sigma_g$ (blue) and from the $1\pi_u$ to the
  $3\sigma_g$ (red) orbital using V$X$Z basis sets. Also 
  shown is spectral decomposition of the FCIQMC ground state (green) as a
  reference . The spectra are not normalized for better display. For the VTZ basis
  set, we also computed $\Sigma_u$-photoemission (PE) and inverse photoemission (IPE) spectra for
  the C$_2^-$ and  C$_2^+$ respectively. All energies and spectra are
  obtained with MaxEnt analytic continuation from 44-48 independent calculations. The experimental values are taken from\cite{M1992}, these are also used to
  attribute singlet and triplet states and the $\pm$-symmetry of the $\Sigma$-states. The zero of the frequency axis is set to
$-75.649\,\mathrm{H}$. The time-step used is $\Delta
t=10^{-3}$ for the VTZ basis set, and
$\Delta t= 10^{-3}$ (green) and $\Delta t = 5\times 10^{-4}$ (red,blue) for the VQZ basis set.
}
\label{fig:panelCarbonDimer}
\end{figure}

Three calculations are shown in Fig.~\ref{fig:panelHubWn} for $U/t=2$, employing 70000, $1.6\times 10^6$ and $17\times10^6$ walkers, with the 
corresponding time contours in the complex plane shown in the inset.  Even
though the resulting spectrum forthe smallest walker number is qualitatively correct, it is
broadened and shifted versus the Lanczos spectrum. Increasing the walker number to $1.6\times 10^6$ gives less severe damping. The 
peaks are still slightly displaced compared to the exact result. For $17\times10^6$ walkers, $\alpha$ is small ($\approx 0.12$) and the spectrum is
 fully resolved with the peaks in their correct positions. The
agreement in the weight distribution also serves as an indicator of the impact
of the walker number. The memory used here is
 270 Mb per processor. This already involves significant performance-memory
 tradeoffs, such that a single replica of this calculation can be run with
 less than 800 Mb total memory, more than a factor of 70 smaller than for 
the exact diagonalization.

Fig.~\ref{fig:panelHubWn}c) shows the PE and IPE spectra for a 24-site cluster with 
22 electrons (24E in Betts' notation\cite{Bett}). This illustrates that 
calculations can be performed for doped systems and for much larger Hilbert spaces 
($\sim 6 \times 10^{12}$) than exact diagonalization. For
the IPE spectrum, the main
quasi-particle peak and the main satellite peak at higher energies are
well-resolved. This spectrum is highly difficult to compute since the initial wave function is highly
multiconfigurational, as it is obtained by eliminating the Hartree-Fock
determinant keeping a very high number of leading determinants with similar
weight. For the PE spectrum, both the main quasi-particle peak as
well as two satellite peaks can be clearly identified. As a comparison, we show 
results using the Hirsch-Fye (HF) method,\cite{HF} based on the Matsubara
formalism for $T=0.2t$.
The HF PE spectrum is consistent with the FCIQMC spectrum, but the peaks are not 
resolved. This is due factors i) and ii) above (performing analytical
continuation from imaginary times and not being able to shift the peak at
-2.5t to 0). The weight
of the IPE spectrum is only $~0.035$ and the relative standard deviation 
about a factor of 25 larger for the part of the Green's function relevant for IPE 
than for the PE relevant part [iii) above].

{\it Application to {\it ab-initio} systems.}
We employ the scheme for {\it ab-initio} systems, namely the carbon atom and the carbon dimer at equilibrium distance. Here, the Hamiltonian is the
molecular Hamiltonian in the Born-Oppenheimer approximation
\begin{equation}
H = \sum_{p,q,\sigma} h^p_{q} c_{p\sigma}^\dagger c^{}_{q\sigma} + \sum_{p,q,r,s,\sigma\tau} V^{rs}_{qp}
c^\dagger_{r\sigma} c^\dagger_{s\tau} c^{}_{p\tau} c^{}_{q\sigma}\,,
\end{equation}
where $h^p_q$ contains the one-body integrals of the \Schrodinger Hamiltonian, and
$V^{rs}_{pq}$ the two-body Coulomb integrals of the
electron-electron interaction. We used the cc-pV$X$Z basis sets with $X$=T,Q (referred to as V$X$Z in the following), 
containing 28 and 54 functions per atom respectively, in the frozen-core approximation. The required Hamiltonian integrals 
were computed over restricted Hartree-Fock orbitals using MOLPRO \cite{MOLPRO-WIREs}. 

For the carbon atom, we show the multiplet structure of the ground state in
Fig.~\ref{fig:panelCarbonStats}, obtained over a trajectory of 1600 a.u. of
time.  

Due to the small system size, we performed the
propagation in pure real-time, with a time-step of $\Delta t = 5\times 10^{-3}$. A small constant
damping with a decay constant of
$3\,\mathrm{mH}$ is applied that has negligible influence on the spectral function, but
reduces the growth of walkers and allows for longer propagation times.
The cation ground state
energy from the ground state computation for the preparation of the initial
state is $E^{N-1}_0=-37.3706\,\mathrm{H}$, which gives an ionization energy of
$420\,\mathrm{mH}$, agreeing reasonably well with the experimental finding of
$413.8\,\mathrm{mH}$ \cite{KW1966}. The inset of Fig.~\ref{fig:panelCarbonStats} shows the
oscillations of the overlap $\Braket{\Psi(0)|\Psi(t)}$ and corresponding
spectra. The resulting excitation energies agree fairly well with experiment.

Next, we consider spectral functions of a prototypical strongly correlated molecule, the carbon dimer  
at equilibrium distance.
To target specific states, we simulate photoabsorption (PA) spectroscopy. 
To do so, the initial $^1\Sigma_g^+$ state is prepared by performing a ground state calculation 
on the neutral carbon dimer using FCIQMC, and then applying the single excitation
operator $c^\dagger_ic^{}_j$ on the resulting walker population. Specifically,
we consider the excitations from $1\pi_u$ to the $3\sigma_g$ and the
excitation from $2\sigma_u$ to $3\sigma_g$. The former couple to $\Pi_u$ states, whilst the latter 
couple to $\Sigma^+_u$ states. Since the excitations generate open-shell determinants, the resulting 
spectra couple to both singlet and triplet states.                                                    

The resulting spectra for the two basis sets                 
are shown in Fig.~\ref{fig:panelCarbonDimer}, we additionally compare to
projector QMC values computed using the excited-state i-FCIQMC method \cite{excited_state_neci} and
using the ground state energies calculated in \cite{BTCA2012} as references. The involved Hilbert spaces contain respectively 
$10^{10}, 10^{12}$ Slater determinants. 
Sharply resolved peaks which correspond to $^3\Pi_u, ^3\Sigma^+_u, 
 ^1\Pi_u, ^1\Sigma^+_u$ could be identified. We also performed photoemission
 and inverse photoemission calculations for the C$_2^-$ and C$_2^+$
 respectively, the resulting energies for the excited states of the neutral
 C$_2$ are listed in Fig.~\ref{fig:panelCarbonDimer}. We find that the inverse
 photoemission spectra feature the lowest stochastic error while the
 photoemission results have a higher error. A rotation of time in the complex plane by an angle of $\alpha$ in the range [0.1,0.2] 
 is applied.
 The dependence of the spectra on the basis set is in line with the known basis-set dependence of 
 relative energies in molecular systems, for example ionisation energies and electron affinities
 from FCIQMC quantum chemical studies  \cite{BA2010,CBA2011,BCTA2011}. The vertical transition energies 
obtained here are larger than the experimentally observed values.  A previous analysis by Holmes et al. \cite{HUS2017} 
of the excited state  potential energy curves shows a significant effect of bond-length variation for the states considered here, 
indicating the likely non-vertical character of the experimental transitions. 

{\it Conclusions.} We have presented an efficient method for solving the time-dependent Schr\"odinger 
equation. We generalize a full configuration interaction Quantum Monte Carlo method to calculations 
for complex times close to the real axis. We then develop a maximum entropy method for analytic
continuation from complex times to real frequency. The method can be used to calculate 
electron spectra. The imaginary component of time strongly limits the numerical effort without 
a strong negative impact on the analytic continuation. We demonstrated that spectra of the Hubbard model 
can be obtained in good agreement with exact Lanczos calculations. We then applied the method to 
{\it ab initio} systems, the C atom and the C$_2$ molecule, and obtained good agreement 
with experiment for excitation energies. The method can be used as cluster solver in embedding 
schemes for solids. It  can also be used to study small systems in strong external 
fields without any assumptions about linear response.
 
%

\clearpage
\section{Appendix}
\subsection{Recap of the FCIQMC method}

The FCIQMC method \cite{BAT2009,BA2010,CBA2010} is a projector quantum Monte Carlo method based on 
the imaginary-time Schr\"odinger equation. 
It has the stationary form
\beq
  {\partial\over \partial \tau} |\Psi\rangle= -(\hat{H}-E_0)|\Psi\rangle = 0,  
\eeq 
with formal solution:
\beq
  |\Psi(\tau) \rangle = e^{-\tau (\hat{H}-E_0)} |\Psi(0) \rangle
\eeq 
which converges (up to a normalization constant) to the ground state $|\Psi_0\rangle$ of $\hat{H}$ in the large $\tau$ limit. 
We define a first-order propagator $\hat{P}$ as
\begin{equation}
  \hat{P}   =   \mathds{1}-\Delta \tau (\hat{H}-S\mathds{1}),   
\end{equation}
where $\Delta\tau$ is a time-step and $S$ an energy shift to control the
walker number.
If $\hat{H}$ has a finite spectral width $W$,
repeated application leads to the ground-state 
\beq
  |\Psi(n \Delta \tau)\rangle  & = & \hat{P}^n |\Psi(0) \rangle \\
  \lim_{n\rightarrow\infty} | \Psi(n \Delta \tau) \rangle  & \propto&  |\Psi_0 \rangle, \nonumber
\eeq 
without a time-step error, if $\Delta\tau$ is smaller than
$\frac{2}{W}$. $|\Psi(\tau)\rangle $ is expressed as a linear combination of a
complete set of basis states ${|D_i\rangle}$ 
\beq
    |\Psi(\tau) \rangle = \sum_i C_i(\tau) | D_i \rangle
\eeq
In FCIQMC, the coefficients  $C_i$  are replaced by 
an ensemble of  positive and negative {\em walkers}:
\beq
   C_i \propto N_i =  \sum_w^{N_w} s_w \delta(i-i_w) \label{deltafunc}
\eeq
where $s_w=\pm 1$ is the sign of the walker $w$, residing on Slater determinant $i_w$. $N_w$ is the number of walkers. 
The walkers evolve according  to stochastic rules
\begin{itemize}
\item A spawning step: a given walker, on $|D_i\rangle$, randomly selects another connected determinant, $|D_j\rangle$ with probability
$p_{gen}(j|i)$. It then attempts to spawn a new walker on $|D_j\rangle$ with probability $p_s=-\Delta \tau  H_{ij} /p_{gen}(j|i)$. 
\item A death/cloning step: A walker on $D_i$ attempts to die with probability $p_d = \Delta \tau(H_{ii}-S) $. 
\end{itemize}
In a following step, walkers with opposite signs cancel each other, which is essential for addressing the sign-problem. 
In the initiator version of the algorithm \cite{CBA2010,CBA2011,BTCA2012} the spawning is restricted. If the target determinant is not occupied by another walker, the spawning  is aborted if 
$|N_i|\le n_a$, where $n_a$ is the initiator parameter. This condition is
crucial for obtaining a smooth convergence without too many walkers.               

For the calculation of reduced density matrices (RDM), we use the replica method \cite{OBCA2014}, 
in which two independent simulations of walkers are propagated and elements of
the RDMs are being calculated by taking products 
involving the two replicas.

The main advantage of the FCIQMC algorithm compared to conventional exact diagonalization is that
the number of walkers needed for convergence is much smaller than the
dimension of the Hilbert space, thereby requiring drastically less
memory. Using this technique, molecular and condensed-matter systems involving
Hilbert spaces of over $10^{20}$ Slater determinants have been computed \cite{Daday2012,Shepherd2012}.

\subsection{Norm conservation}
Compared to the pure imaginary time evolution, the complex exponential in the
real-time formulation does not cause an exponential decay of contributions
from excited states, but instead gives a complex phase to the walkers, which
requires the use of both real and imaginary walkers for each determinant. Here, real and
imaginary populations are only coupled via the stochastic application of the
first-order expanded propagator
\begin{equation}
\hat{U}_1(t) = 1 - \mathrm{i}\hat{H}t\,.
\end{equation}
The annihilation step is performed separately for each of the populations.

The direct use of $\hat{U}_1$ in the time propagation leads to an exponentially increasing  
wave function, and therefore severely violates norm-conservation of unitary dynamics. 
This can be seen by considering the time evolution of a wave function $\Psi=\Psi_0$ that is already an eigenstate of the 
Hamiltonian with energy $E$.
The exact solution is 
\begin{equation}
  |\Psi(t)\rangle = e^{-\mathrm{i}Et} |\Psi_0 \rangle   
\end{equation}
According to the first-order propagator, after $n$ application of $\hat{U}_1$
we obtain:
\begin{equation}
|\Psi(t_n)\rangle = (1 - \mathrm{i} E \Delta t)^n |\Psi_0 \rangle  \,,
\end{equation}
with $t_n = n \Delta t$, we obtain:
\begin{eqnarray}\label{eq:rungekutta}
  {\rm ln}{\Psi(t_n)\over \Psi_0}  =  n {\rm ln} (1 - \mathrm{i} E \Delta t) 
   \approx & -\mathrm{i} E t_n + {E^2 t_n  \Delta t\over 2}                              
\end{eqnarray}
so that:
\begin{equation}\label{eq:rungekutta1}
|\Psi(t_n)\rangle = e^{-\mathrm{i} E t_n} e^{E^2 t_n \Delta t/2 } |\Psi_0 \rangle    
\end{equation}
which is exponentially growing in time, with an exponent ${\cal O}(\Delta t)$.                                            
This is a direct consequence of working with real time, which introduces a growing exponential factor 
in Eq.~(\ref{eq:rungekutta1}).

This problem can be greatly suppressed using a second-order short-time propagator. 
Defining:
\beq
\hat{U}_2(\Delta t) = \mathds{1} - \mathrm{i}\Delta t\hat{H} - \frac{1}{2}(\Delta t)^2 H^2
\eeq

The time evolution is implemented using a second-order Runge-Kutta
algorithm, which decomposes $\hat{U}_2$ into two steps:
\begin{equation}
\hat{U}_2(\Delta t) = \mathds{1} + \left(\hat{U}_1(\Delta t) -
  \mathds{1}\right)\hat{U}_1\left(\frac{\Delta t}{2}\right)\,.
\end{equation}
The second order propagator is applied by first applying
$\hat{U}_1\left(\frac{\Delta t}{2}\right)$ to the wavefunction, followed by applying
$\left(\hat{U}_1(\Delta t) - \mathds{1}\right)$ to the result and finally adding
the resulting wavefunction to the original one. In this way, $\hat{H}^2$ is not
explicitly applied, which is highly advantageous for the efficiency of the method.

We now have after $n$ repetitions of $\hat{U}_2$: 
\beq
|\Psi(t_n)\rangle = [1 - i E \Delta t - \frac{(E \Delta t) ^2}{2} ]^n |\Psi_0  \rangle  
\eeq
resulting in: 
\begin{eqnarray}\label{eq:second}
  \ln {\Psi(t_n)\over \Psi_0} & = & n \ln [1 - \mathrm{i} E \Delta t -  \frac{(E \Delta t)^2}{2} ] \\
  & \approx&  -\mathrm{i} E t_n  - \mathrm{i} \frac{  E^3 t_n  (\Delta t)^2}{6} + 
      \frac{E^4 t_n  (\Delta t)^3}{8}        \nonumber           
\end{eqnarray}
\beq
|  \Psi(t_n)\rangle = e^{-\mathrm{i} E t_n} e^{-\mathrm{i} E^3 t_n (\Delta t)^2/6} e^{E^4 t_n (\Delta )^3/8} |\Psi_0 \rangle  \nonumber                
\eeq
i.e. in this formulation the norm-violating factor grows only as ${\cal O}(\Delta t^3)$.       
It is possible to reduce further the scaling of norm-violation 
by employing a 4-th order propagator, but we found that improvements are typically masked           
by much larger stochastic errors.

\begin{figure}[t]
\includegraphics[width=0.8\columnwidth]{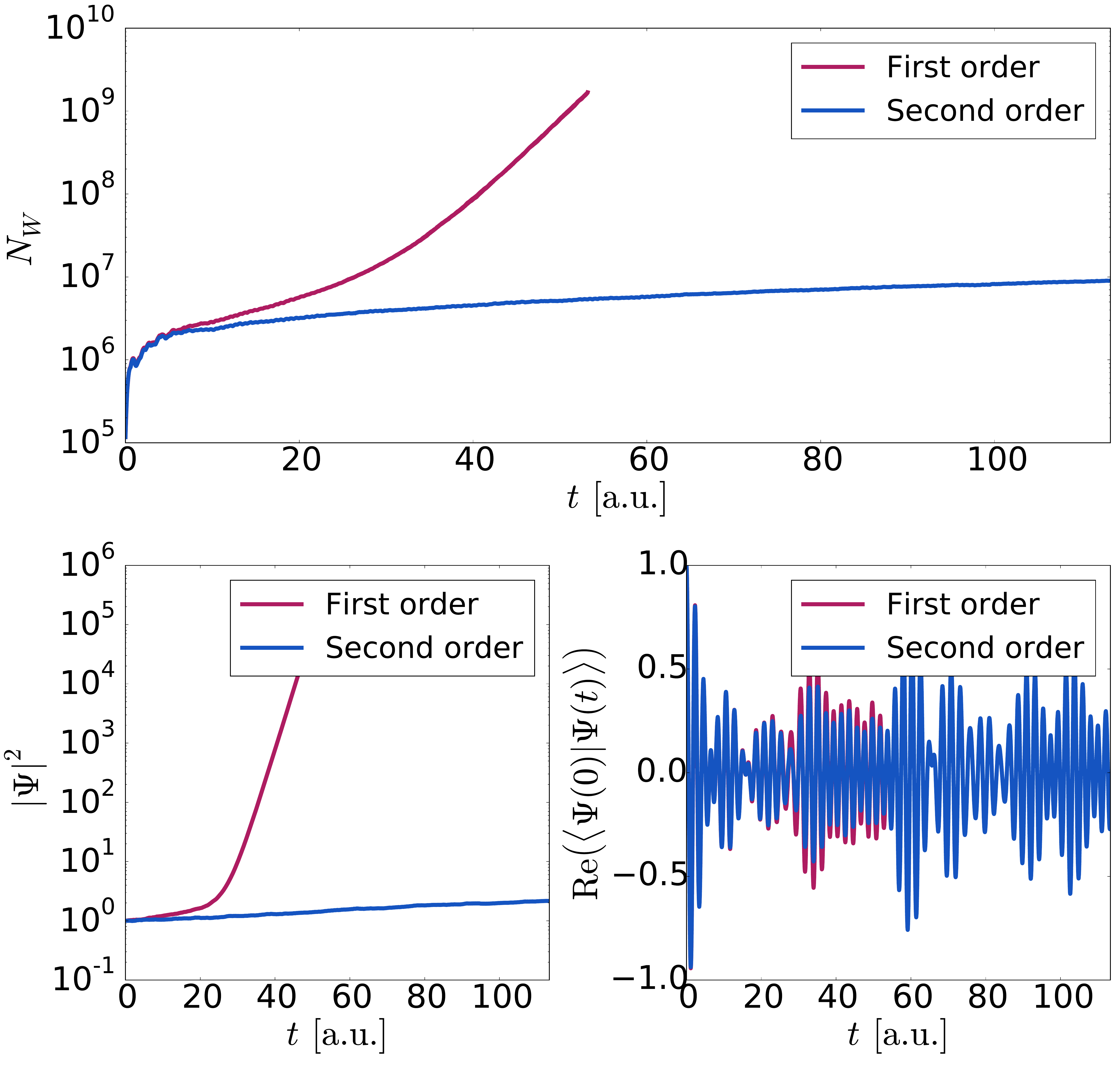}
\caption{Number of walkers $N_W$, norm of the wave function, and Green's function 
for the 10-site Hubbard model,  $k=(0,0)$ and  $U/t=1$, using the stochastic algorithm
 and the first and second order Runge-Kutta methods.}
\label{fig:firstOrder}
\end{figure}

Fig.~\ref{fig:firstOrder} compares the first and second order expansions in the Rung-Kutta method.
The figure illustrates how the number of walkers and the norm rapidly increase
in the first order expansion. The first order overlap $\langle \Psi_i^{\pm}(0)|\Psi^{\pm}_i(t)\rangle$
is substantially more accurate than the norm, but still not satisfactory.

\begin{figure*}[t]
\includegraphics[width=0.99\textwidth]{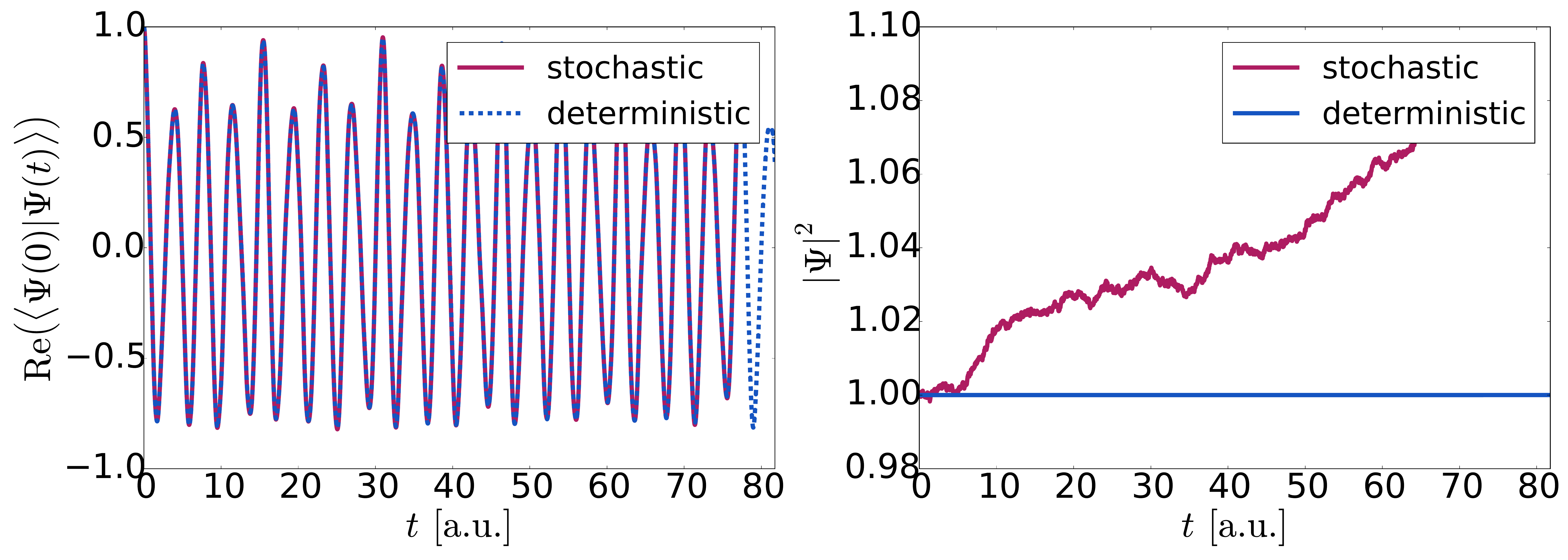}

\caption{Green's function (left) and norm (right) of the wave function over time  
using the second-order algorithm for the two-dimensional 10-site Hubbard model, 
$k=(0,0)$ and $U/t=1$. Both were calculated using both stochastic and deterministic algorithms.}
\label{fig:normError}
\end{figure*}

Fig.~\ref{fig:normError} (right part) compares deterministic\cite{PHC2012,BSKS2015} and 
stochastic calculations of time evolutions of the norm $\langle \Psi(t)|\Psi(t)\rangle$
to second order.  The deterministic calculation only contains the errors of the second 
order Runge-Kutta, and it is very accurate over this time scale.  The stochastic calculation 
introduces substantial errors in the norm, e.g., due to excitations to high-lying states. 
In the overlap $\langle \Psi_j^{\pm}(0)|\Psi^{\pm}_i(t)\rangle$ these stochastic errors tend 
to cancel (see left part of Fig.~\ref{fig:normError}) for two reasons. Many of the stochastically 
excited states have little or no weight in the initial state and therefore give little 
or no contribution to the overlap. Furthermore, the stochastic errors due to the time evolution 
enter linearly in $\langle \Psi_j^{\pm}(0)|\Psi^{\pm}_i(t)\rangle$ and therefore tend to cancel. 
This is crucial for the accuracy of the method. 
We could alternatively have calculated $\langle\Psi_j^{\pm}(t/2)| \Psi^{\pm}_i(t/2)\rangle$, but 
in this case the stochastic errors are much larger, since the the two arguments above do not apply.   

Even though a symplectic integrator such as the Verlet method \cite{Verlet} could in
principle yield smaller discretization errors, we find that stochastic errors
play a much larger role, making it unfeasible compared to the
Runge-Kutta integrator.

\subsection{Computation of Green's functions and optical absorption}
Here, we provide some more details about the calculation of the Green's function.

We assume that the ground-state $|\Psi_0^N\rangle$ for $N$ electrons has been calculated.
We then want to calculate the Green's function
\begin{equation}\label{eq:g0}
G_{ij}(t)=-i\langle\Psi_0^N|T\lbrace c_i(t)c_j^{\dagger}\rbrace|\Psi_0^N\rangle,
\end{equation}
where $T$ is the time-ordering operator, $c_i$ is the annihilation operator for an electron 
with quantum numbers $i$ (including spin) and $c_i(t)={\rm exp}(i\hat Ht)c_i{\rm exp}(-i\hat Ht)$. 
For $t<0$ ($t>0$) this corresponds to (inverse) photoemission. For photoemission we make 
a variable substitution $t \to -t$. Then both photoemission and inverse photoemission 
correspond to positive time propagation, but there is now an extra minus sign in the 
Schr\"odinger equation for photoemission. We then consider the initial state
\begin{equation}\label{eq:g1}
|\Psi^{\pm}_i(0)\rangle=c_i^{\pm}|\Psi_0^N\rangle,
\end{equation}
where lower (upper) sign indicates (inverse) photoemission and $c_i^{+}=c_i^{\dagger}$ and $c_i^{-}=c_i$.
We solve the Schr\"odinger equation
\begin{equation}\label{eq:g2}
i{d\over dt}|\Psi_i^{\pm}(t)\rangle=\pm [ e^{\mp i\alpha(t)}(\hat H-E_0^N\mp \mu)]|\Psi_i^{\pm}(t)\rangle,
\end{equation}
Here $\alpha(t)$ defines the path through the complex time plane. 
$\alpha(t)\equiv 0$ ($\pi/2$)) corresponds to integration along the real (imaginary) time axis. 
The formal solution can be written as 
\begin{eqnarray}\label{eq:g3}
&&|\Psi_i^{\pm}(t)\rangle \\
&&={\rm exp}[ \mp i\int_0^t dt'e^{\mp i\alpha(t')}(\hat H-E_0^N\mp \mu)] |\Psi_i^{\pm}(0)\rangle \nonumber
\end{eqnarray}
We take the overlap to the state $\langle \Psi^{\pm}_j(0)|$ and expand this in a complete set of states
$|\Psi_n^{N \pm 1}\rangle$.
\begin{eqnarray}\label{eq:g4}
&& \langle \Psi_j^{\pm}(0)|\Psi_i^{\pm}(t)\rangle   \\
&&=\sum_n\langle \Psi_0^N|c^{\mp}_j|\Psi_n^{N\pm 1}\rangle   \langle \Psi_n^{N\pm 1}|c^{\pm}_i|\Psi_0^N\rangle  \nonumber  \\
&&\times {\rm exp}\lbrace \mp i\int_0^tdt' {\rm exp}[\mp i\alpha(t')] [E_n^{N\pm 1}-E_0^N\mp \mu]\rbrace  \nonumber \\
&&=\int d\omega A^{\pm}_{ji}(\omega){\rm exp}\lbrace -i\int_0^t dt' {\rm exp}[\mp i\alpha(t')]\omega\rbrace  \nonumber
\end{eqnarray}
Here we have introduced the spectral functions
\begin{eqnarray}\label{eq:g5}
&& A^{\pm}_{ji}=\sum_n \langle \Psi_0^N|c^{\mp}_j|\Psi_n^{N\pm 1}\rangle    \\
&&\times \langle \Psi_n^{N\pm 1}|c^{\pm}_i|\Psi_0^N\rangle \delta[\omega\mp E_n^{N\pm 1}\pm E_0^N+\mu] \nonumber
\end{eqnarray}
We finally introduce the spectral function
\begin{equation}\label{eq:g6}
A_{ij}(\omega)=A_{ij}^{+}(\omega)+A_{ij}^{-}(\omega),
\end{equation}
where we have used conventions that negative  (positive) frequencies correspond to (inverse) photoemission.
Large (small) values of $|\omega|$ correspond to excited states with large (small) excitation energy.
In a similar way we can calculate optical conductivity, by applying a current operator
to the $N$-particle state and propagating this in time.

\begin{figure*}
\includegraphics[width=\columnwidth]{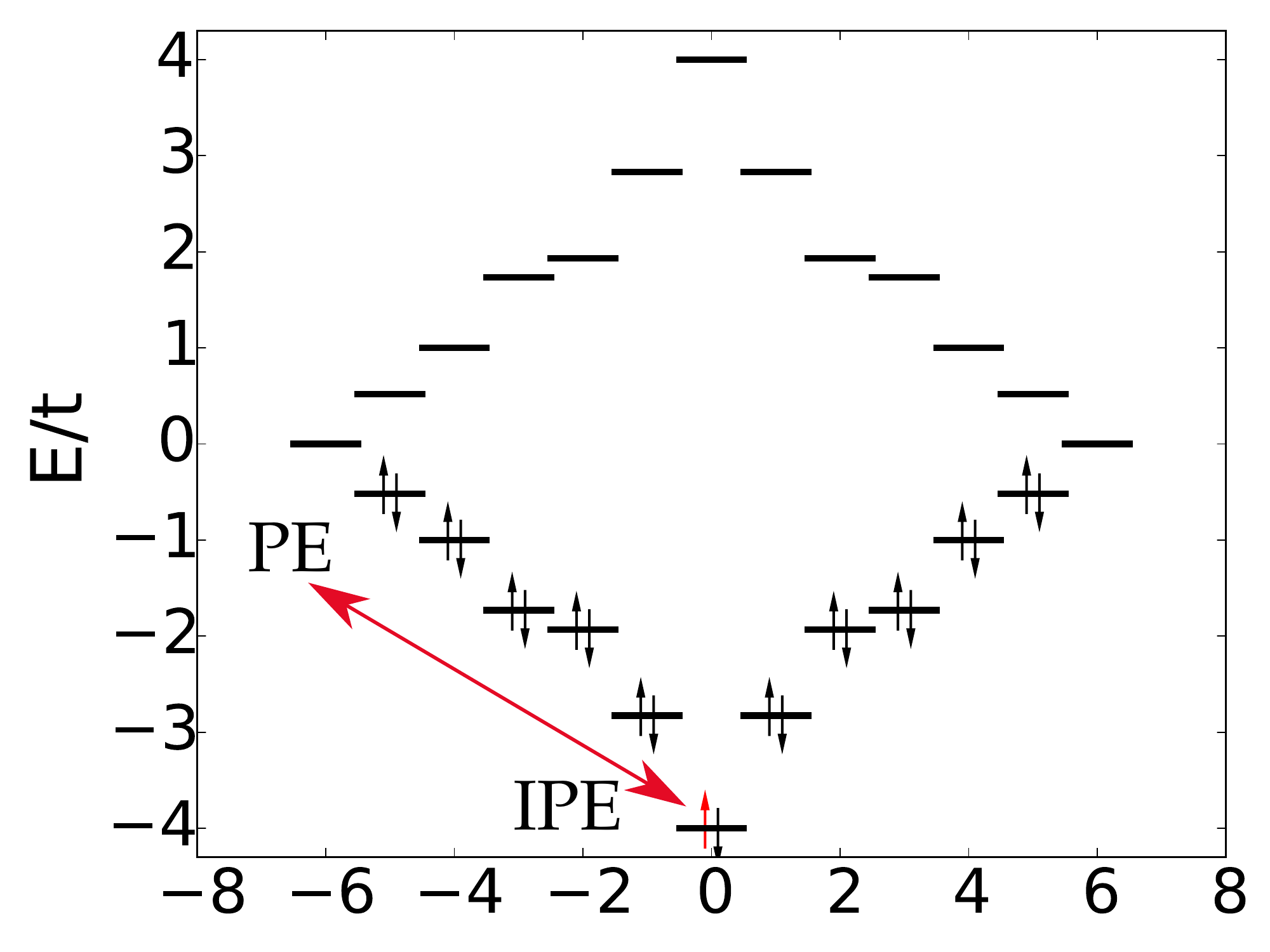}
\caption{Energy levels of the non-interacting $24$-site Hubbard model with
  lattice vectors (3,3) and (-5,3). The (inverse) photoemission spectrum for $k=(0,0)$ is
  obtained by removing (adding) the electron marked in red. While removing an
  electron with $k=(0,0)$ keeps the determinant with highest weight and
  therefore creates an initial state with a unique high-weight leading
  determinant, this is not the case for the inverse photoemission. As
  $k=(0,0)$ is doubly occupied in the reference determinant of the ground
  state, the latter does not appear in the initial state and we start from a
  enormously correlated state with a high number of determinants with
  comparable weight.}
\label{fig:dispersion_ipe_pe}
\end{figure*}

The targeted spectral function then dictates the structure of the initial
wavefunction, and thereby also the level of correlation present in the initial
state.
As $\Psi_0^N$ is taken from a previous FCIQMC calculation, the initial state
is obtained from a stochastic sample of the true ground-state. Therefore, multiple
independent samples of $\Psi_0^N$ are taken, and the Green's function is
computed from the overlap of the initial state of one sample with the
time-evolution of another, since a Green's function from only a single sample
is quadratic in the initial state and is hence potentially biased. We find
that such a bias is problematic only for the most correlated initial states,
like the inverse photoemission for the $24$-site Hubbard model as in
Fig. ~\ref{fig:dispersion_ipe_pe}, but using a Green's function obtained from
a single sample should be avoided nevertheless.

\subsection{Complex time contour}

We use a time-dependent angle $\alpha(t)$, which is adjusted so that the number 
of walkers do not appreciably exceed a preset value. This is done in a similar way 
as the walker number control in the projective algorithm. We prescribe 
an initial value $\alpha(t=0) = \alpha_0$, typically $\alpha_0=0$. Once the
walker number exceeds a threshold value $N_\mathrm{target}$, we start to adjust $\alpha$ every $B$
steps as 
\begin{equation}
\alpha (t + B\Delta t) = \alpha (t) + \xi\, \mathrm{arctan}\left(\frac{N_W\left(t+B\Delta
      t\right)}{N_W\left(t\right)} -1 \right) \,.
\end{equation}
Here, $N_w(t)$ is the number of walkers at time $t$ and $\xi \sim 0.1\, - \, 1.0$ is a damping
parameter. Using this heuristic approach, the value of alpha is iteratively
updated to counter changes in the walker number. We use the $\mathrm{arctan}$ function
to map changes in walker number to changes in an angle, but for sufficiently
small $B \approx 10 $, we do not expect the exact choice of the function used for
this mapping to have
an impact. Using this technique, the value of $\alpha$ is increased during the
time evolution as the walker number increases, which in turn damps the walker
number growth, eventually leading to an equilibration of both the value of $\alpha$
and the number of walkers. However, depending on the chosen parameters $\xi$
and $B$, even in equilibrium, the value of $\alpha$ can be subject to rapid
fluctuations around the average value due to short-time fluctuations in the
number of walkers. This has no notable impact on the contour, however. The equilibrium value of $\alpha$ is then typically $\sim
0.05\,-\,0.25$ for the studied systems, except for the $24$-site Hubbard model
with an equilibrium value of $\alpha \sim 0.45$.
Increasing the walker threshold value $N_\mathrm{target}$ tends to decrease $\alpha$. 
 
\subsubsection{Walker number dependence}

The walker number impacts the time-evolution in two ways. The first is the
influence on the adaptation of $\alpha$, as increasing the walker number for a
fixed initial number of walkers lowers the required values of $\alpha$ for a
stable calculation with a constant walker number. The control mechanisms for
adjusting the walker number here are setting the initial value $\alpha_0$
and/or a minimum walker number which has to be reached before the value of
$\alpha$ is changed. In particular only adjusting $\alpha$ once a given
number of walkers is reached allows for targeting specific walker numbers,
similar to the variable shift mode in the projected algorithm, although the
walker number equilibration is typically slower. The values
$\alpha$ obtains in this procedure decrease as the targeted walker number is
increased, while increasing $\alpha_0$ unsurprisingly decreases the number of
walkers used. 

\begin{figure*}[t]
\includegraphics[width=0.9\textwidth]{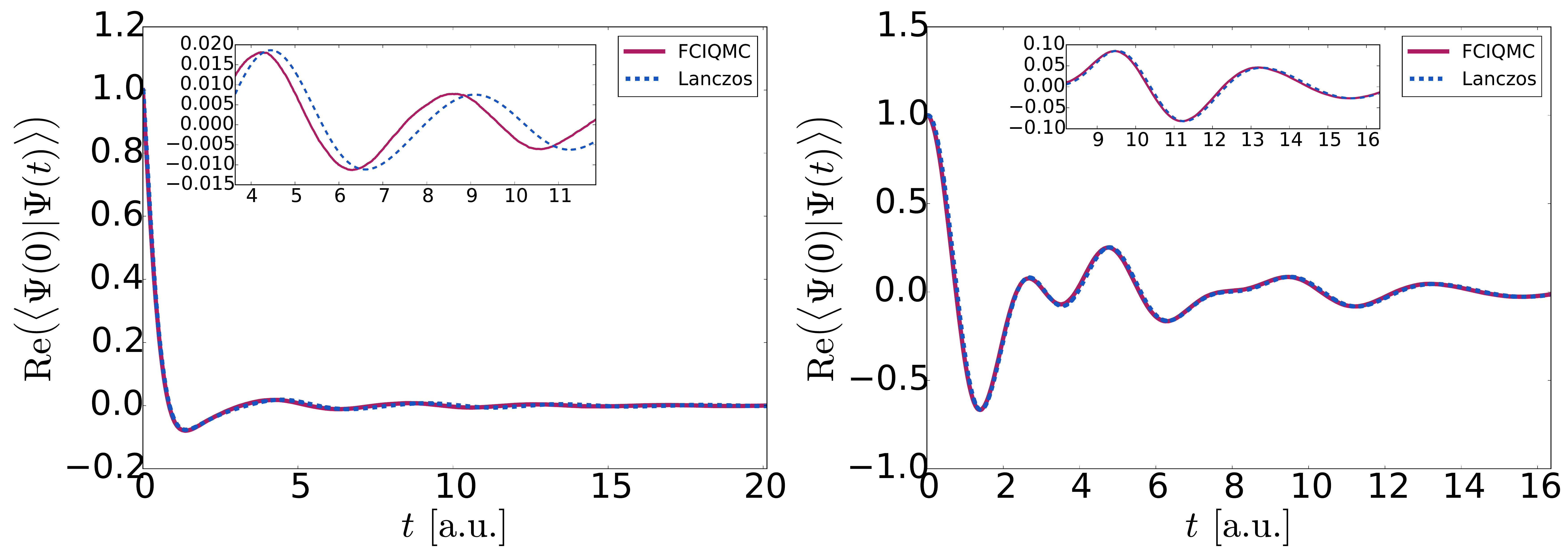}
\caption{Photoemission Green's function for $k=(0,0)$ for the Hubbard model
  with an 18-site cluster at $U/t=2$ obtained with FCIQMC and Lanczos with a)
  70000 and b) 17 million walkers, showing a bias in the Green's function due to under-sampling for the
  smaller walker number. }
\label{fig:24siteWN}
\end{figure*}

The second effect is a bias in the Green's function itself as shown in figure ~\ref{fig:24siteWN}.

\subsection{Chemical potential shift}

Typically we are particularly interested in the spectrum relatively close to the
chemical potential (within several eV). We can emphasize these states by using the 
flexibility of the present method. Thus we study the spectra for each ${\bf k}$ at 
a time and photoemission and inverse photoemission separately. We then have the 
freedom to choose the chemical potential as $E_0^N-E_0^{N-1}({\bf k})\le \mu \le 
E_0^{N+1}({\bf k})-E_0^N$ in the spectral calculation, where $E_0^{M}({\bf k})$ 
is the lowest $M$-electron state with the wave vector ${\bf k}$. Lowering (increasing) 
$\mu$ for (inverse) photoemission leads to a slower decay of the Green's function 
for a given $\alpha(t)$. The shift increases the weight of all states. To keep the 
number of walkers fixed,  $\alpha(t)$ is then increased. This suppresses high-lying 
states (far from $\mu$) more than low-lying states, enhancing the relative weight 
of low-lying states, as the suppression scales with energy. The result is that 
low-lying states contribute to the Green's function over a longer time, and it 
then becomes easier to extract the information about these states. This should 
then also improve the signal to noise ratio for low-lying states. Fig.~\ref{fig:maxent} 
(e.g., for $\alpha_0=\pi/4$ or 0.2) illustrates how structures close to $\mu$ 
are described more accurately.

We can use 
\begin{equation}\label{eq:g7}
\mu=\left\{ \begin{array}{ll}
E_0^{N+1}({\bf k})-E_0^N &   {\rm inverse \ photoemission} \\
E_0^N-E_0^{N-1}({\bf k}) &  {\rm photoemission}
\end{array}\right.
\end{equation}
In this way the contribution to the spectrum from $|\Psi_0^{N\pm1}({\bf k})\rangle$ 
is not damped by $\alpha(t)$, and its contribution to the spectrum is therefore well 
described.

Sometimes the lowest states of the $(N\pm 1)$-system with a given ${\bf k}$ have 
little or no weight in the spectrum of interest and it may then be favorable to 
reduce (increase) $\mu$ even more for (inverse) photoemission. Eventually, however, 
these states obtain weight due to statistical noise and then grow exponentially. 
The shift of $\mu$ should therefore not be too large.

The Matsubara formalism has often been used to study the Mott metal-insulator 
transition or the formation of a pseudo gap. Then the (angular integrated)
spectrum  at $\mu$ is of particular interest, and the Matsubara formalism 
provides very useful information. However, we are often also interested in angular resolved spectra, where for a given ${\bf k}$ the leading peak may be located
well away from $\mu$. Then the separate treatment of each ${\bf k}$ in the
present formalism, and the related possibility to shift the spectrum,
becomes particularly important. Satellites are also often of interest, 
and then the use of a relatively small $\alpha(t)$ in the FCIQMC is
of great advantage.

In the Matsubara formalism the photoemission and inverse photoemission spectra
are treated simultaneously. In the ${\bf k}$-resolved case the relative weights,
and thereby the relative standard deviations, may be very different. The present
separate treatment of the two spectra then becomes an important advantage, since 
the relative standard deviations are comparable for the two spectra.

\subsection{The initiator approximation}

We make use of the initiator version of FCIQMC \cite{CBA2010, CBA2011} which
is commonly used in the projective algorithm. This limits the possibilities
for walkers to spawn to unoccupied determinants and thereby prevents sign
errors from proliferating. The adaptation made is, that spawns onto
unoccupied determinants are only accepted if they either came from a
determinant exceeding a certain threshold occupation or if another spawn
onto the same determinant occurred in the same iteration. 

In contrast to the projective algorithm, the threshold value itself is not
very significant for the purpose of Green's function calculation, as the initial
wave function already has a high number of determinants populated, and only
their population will enter the Green's function. Also, the event of two
spawns occurring onto the same determinant is common, limiting the influence of
the threshold further. 

\begin{figure*}[t]
\includegraphics[width=0.99\textwidth]{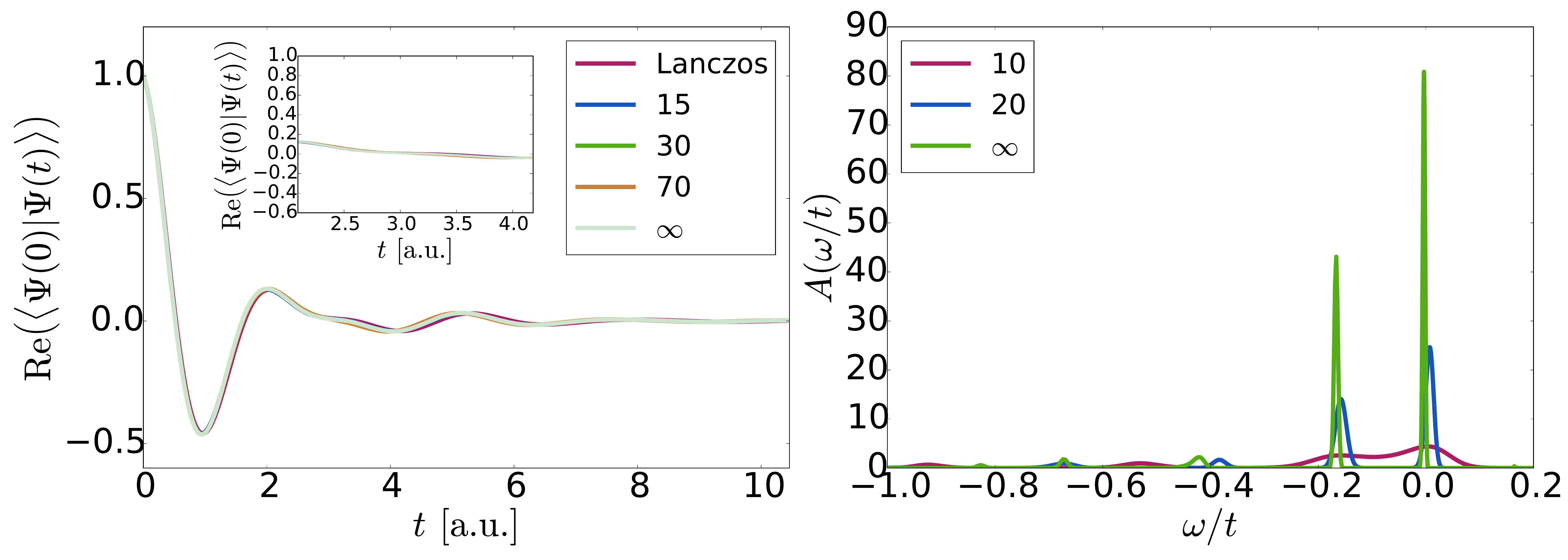}
\caption{(a) Green's function of the $U/t=2$ 18-site Hubbard model for fixed
  $\alpha=0.2$ for different initiator thresholds and without any initiators ($\infty$),
  allowing only double spawns to populate new determinants, and
  as obtained using Lanczos. (b) Photo absorption spectra of $C_2$ in the
  cc-pVTZ basis set for different thresholds and without initiators.
Large values of $\alpha$ were used for the smaller thresholds, leading
to broader spectra.}
\label{fig:init_compare}
\end{figure*}

It can then be highly beneficial to either pick a high threshold, or entirely
disable the possibility to spawn onto unoccupied determinants by single spawns 
and require two spawns to populate a new determinant. Fig.   
~\ref{fig:init_compare} shows the effect of the threshold onto the Green's
function and the spectral function for exemplary cases. The effect on the
Green's function is minor. For the $C_2$ molecule, the high-energy part of the
spectrum exhibits some sensitivity, whereas the low-energy part notices only a
constant shift which does not enter energy differences. 

\subsection{Maximum entropy}
\label{sec:maxent}
\begin{figure*}[t!]
\includegraphics[width=\textwidth]{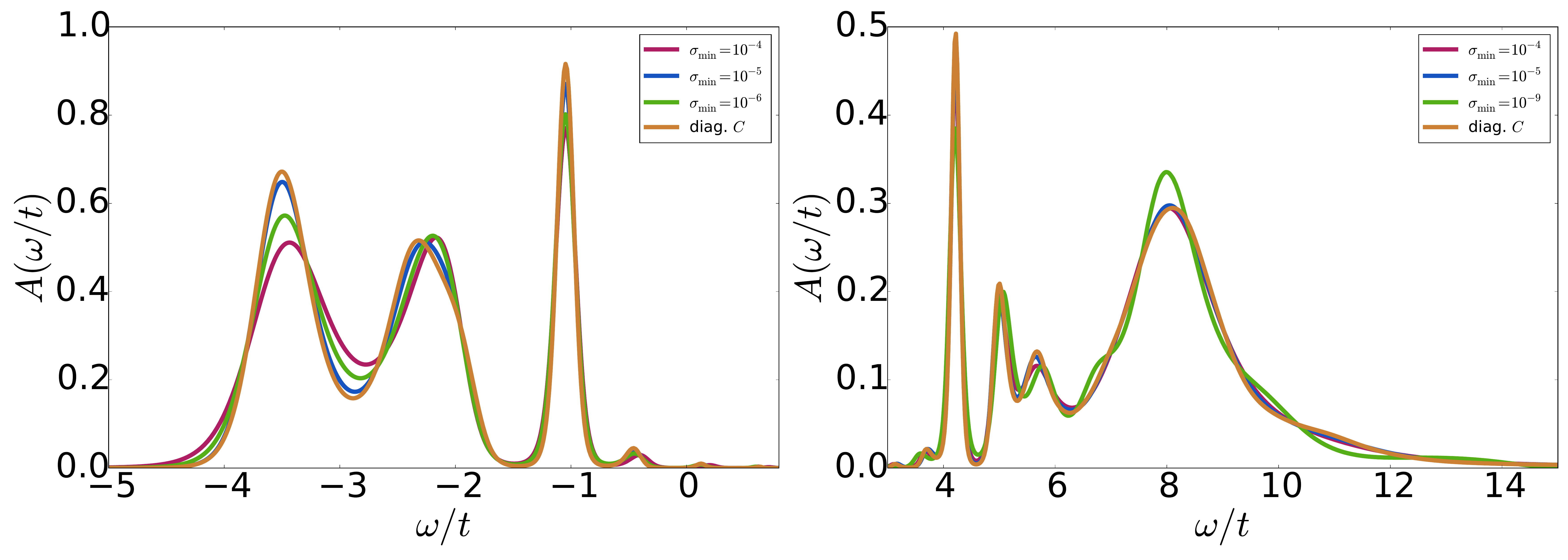}
\caption{Spectra obtained using maximum entropy obtained using different
  values of the cutoff $\sigma_\text{min}$ for the photoemission (left) and the inverse photoemission of the 22-electron 24-site (right) Hubbard model at
  $U/t=4$. While the photoemission spectrum shows sensitivity to the cutoff, the
inverse photoemission spectrum does not as long as the covariance matrix is
non-singular. For $\sigma_\text{min}= 10^{-9}$, this is no longer the case here
and the analytic continuation is ill-defined, leading to deviations in the
spectrum. For comparison, the spectrum obtained by partitioning the data in 6
batches, assuming a diagonal $C$ and averaging over the spectra obtained from
each batch is also shown.}
\label{fig:sigmamin}
\end{figure*}

The maximum entropy method \cite{maxent,Jarrell} for calculating spectral functions is 
often applied together with the finite temperature Matsubara formalism, where the spectral data 
are then analytically continued from the imaginary to the real axis. Here we develop a 
formalism for analytic continuation from an arbitrary path in the complex
plane to the real axis, using the (inverse) photoemission spectrum as an example. 
The spectrum $A_{ij}(\omega)$ is related to the solution of the Schr\"odinger equation via
\begin{equation}\label{eq:max3}
g_k=\sum_l K_{kl}a_l, 
\end{equation}
where $g_k=\langle \Psi_i^{\pm}(0)|\Psi_j^{\pm}(t_k)\rangle$, $a_l=A_{ij}^{\pm}(\omega_l)$ and                                                           
\begin{equation}\label{eq:max1}
K_{kl}={\rm exp}\lbrace -i\int_0^{t_k}dt e^{\mp i\alpha(t)} \omega_l \rbrace f_l,
\end{equation}
where $f_l$ is a weight factor for the $\omega$ integration and the lower (upper) sign         
refers to (inverse) photoemission. The indices $i$ and $j$ have been dropped for simplicity.
We introduce the average $\bar g_k$ over many samples of $g$ and define the deviation $\chi$ of 
a spectral function $a$ giving $g$ from $\bar g$ as
\begin{equation}\label{eq:8}
\chi^2=\sum_{k=1}^L\sum_{k=1}^L({\bar g}_{k}-g_{k})^{*}[C^{-1}]_{kl}({\bar g}_{l}-g_{l}).
\end{equation}
where the sums run over the $L$ values of $g_k$ and $C$ is the covariance
matrix \cite{maxent,Jarrell} of the samples of $g$. To obtain a regular
expression for $\chi^2$, it is important to have a non-singular covariance
matrix $C$, as the inverse $C^{-1}$ is required to calculate
$\chi^2$. If few samples are used, $C$ may be ill-behaved. We have then imposed
a minimum value, $\sigma_\text{min} \approx 10^{-4}\dots 10^{-6}$, on the
diagonal entries of $C$. While this allows for regularizing $C$, it also
assumes the data to be more noisy than it actually is and hence can
affect the details of the spectra as illustrated in Fig.~\ref{fig:sigmamin}. 
Alternatively, we have split the data in batches and assumed a diagonal $C$
for each batch. This assumption can overemphasize noise, which tends to be 
compensated by averaging over batches.

\begin{figure}[t!]
\includegraphics[width=\columnwidth]{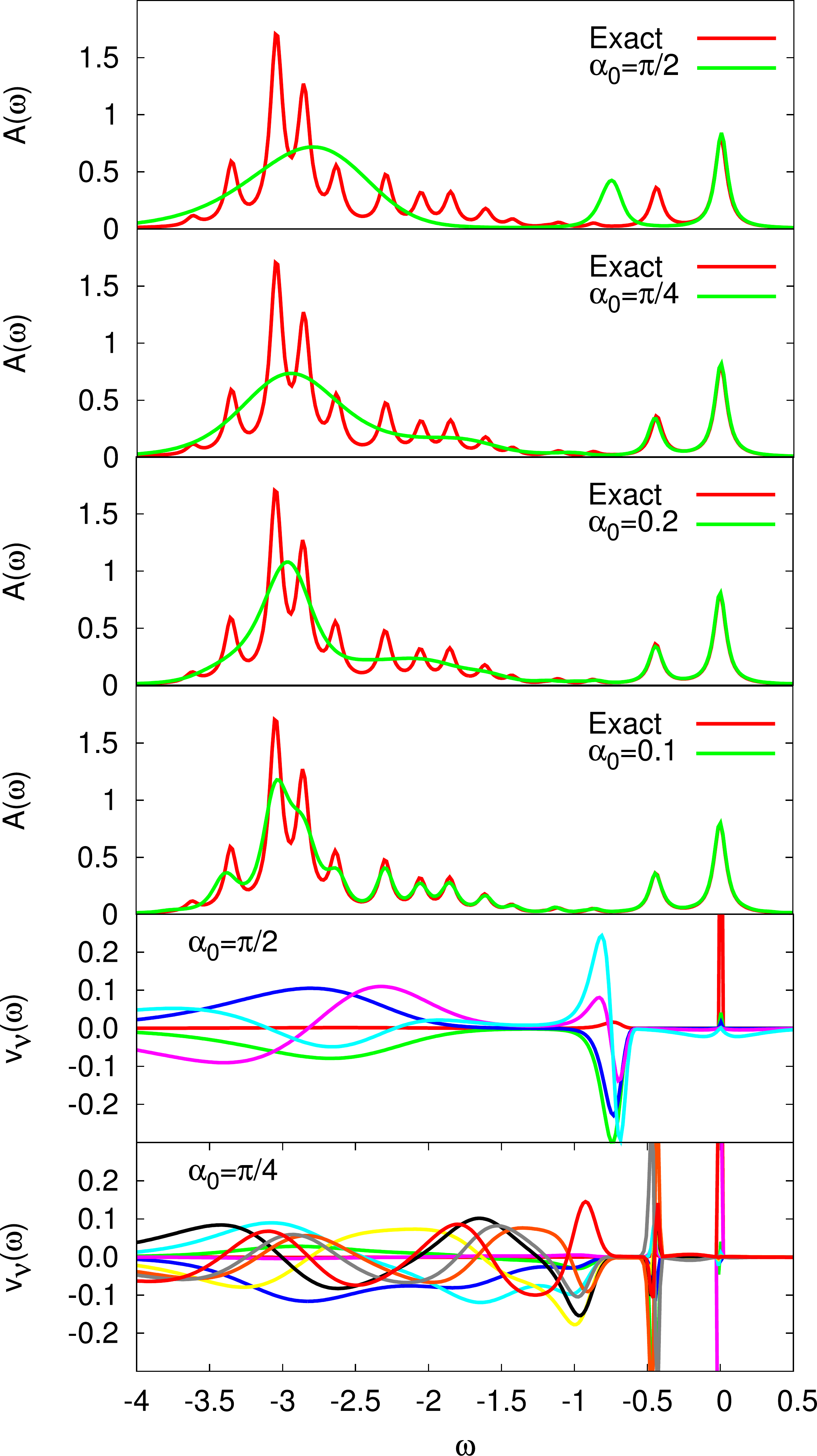}
\caption{Photoemission spectrum for the Hubbard model with 18 sites, $U/t=8$ for different functions $\alpha(t)\equiv \alpha_0$ (top four figures).
The bottom two figure shows the  basis functions $|\nu\rangle$ with 
$\varepsilon_{\nu}>5$ for $\alpha_0=\pi/2$ and $\pi/4$.
The chemical potential is chosen so that one peak is at $\omega=0$.
The data have Gaussian noise with a relative standard deviation 
of about 10$^{-2}$. The spectra have been given a Lorentzian broadening 
with FWHM=0.1.
}
\label{fig:maxent}
\end{figure}

We also introduce the entropy $S$ 
\begin{equation}\label{eq:11}
S=\sum_{i=1}^L[a_i-m_i-a_i{\rm ln}(a_i/m_i)]f_i,
\end{equation}
where $m_i$ is default function providing a guess for $A(\omega)$.
We minimize $\chi-\gamma S$, where $\gamma$ determines the importance of the entropy.
The most probable value of $\gamma$ is chosen \cite{maxent,Jarrell}.
This leads to a system of nonlinear equations.
This system is solved iteratively, by linearizing the equations around successive approximations $a^{(m)}_i$.
We introduce $a_i^{(m+1)}=a_i^{(m)}+\delta a_i^{(m+1)}$ and solve 
\begin{eqnarray}\label{eq:16}
&&\sum_j {\rm Re} [K^{\dagger}C^{-1}]_{kj}{\bar g}_j-\sum_{jl} {\rm Re} [K^{\dagger}C^{-1}K]_{kl}a^{(m)}_l \nonumber \\
&& -\gamma{\rm ln} { a_k^{(m)}\over m_k} =\sum_{l} \Lambda_{kl}(a_k^{(m)})\delta a_l^{(m+1)}. 
\end{eqnarray}
where 
\begin{equation}\label{eq:maxent10} 
\Lambda_{kl}(a_k^{(m)})=\lbrace {f_k\gamma\over a_k^{(m)}}\delta_{kl}+ {\rm Re} [K^{\dagger}C^{-1}K]_{kl} \rbrace
\end{equation}
Fig.~\ref{fig:maxent} show results for the Hubbard model with four different $\alpha(t)\equiv \alpha_0$.
The spectrum was obtained from exact diagonalization, transformed to complex $t$ and Gaussian noise was added.
The spectrum was then transformed back to real frequencies using maximum entropy and compared with the exact result.
For data on the imaginary axis ($\alpha_0=\pi/2$), the $\omega=0$ peak is accurately described, while the other structures 
are approximated by two peaks. For data close to the real axis ($\alpha_0=0.1$) almost all structures are reproduced.

To understand what accuracy can be obtained, we expanded the work in Ref. \onlinecite{analyticcontinuation} and introduce the eigenvectors $|\nu\rangle$ 
and eigenvalues $\varepsilon_{\nu}$ of $\Lambda$. We expand the differences $\delta a=a-a_{\rm exact}$,  $\delta m=m-a_{\rm exact}$
and the stochastic error in $\bar g$ in the eigenvectors $|\nu\rangle$ and obtain coefficient $\delta a_{\nu}$, $\delta m_{\nu}$
and $\delta g_{\nu}$, satisfying    
\begin{equation}\label{eq:maxent11}
\delta a_{\nu}={1\over \varepsilon_{\nu}}(\delta g_{\nu}+\delta m_{\nu}).
\end{equation}
Typically there are several very large $\varepsilon_{\nu}$.
The corresponding components of $a$ are then very accurately described. Other eigenvalues 
are approximately unity, and the corresponding $\delta a_{\nu}$ cannot be
trusted.                        

The bottom of Fig.~\ref{fig:maxent} shows the eigenvectors for $\alpha_0=\pi/2$ and $\pi/4$. 
The eigenvalues for $\alpha_0=1$ are $2\times 10^6$, $3 \times 10^4$, $5 \times 10^3$, 73, 5.   
The components of $A(\omega)$ corresponding to the first four or five eigenvectors are then described very well.
These eigenvectors do not have enough nodes to describe details away from $\omega=0$. As $\alpha_0$ is reduced 
the number of eigenvalues larger than 5 increases from 10 ($\alpha_0=\pi/4$) or 20 ($\alpha_0=0.2$) to about 40 ($\alpha_0=0.1$).
Correspondingly, more and more details of the spectrum can be described.
The $|\nu\rangle$ and $\varepsilon_{\nu}$ help  us judge which details of $A(\omega)$ can be described
and which cannot. 

\subsection{Additional Data}
\begin{figure}[t!]
\includegraphics[width=\columnwidth]{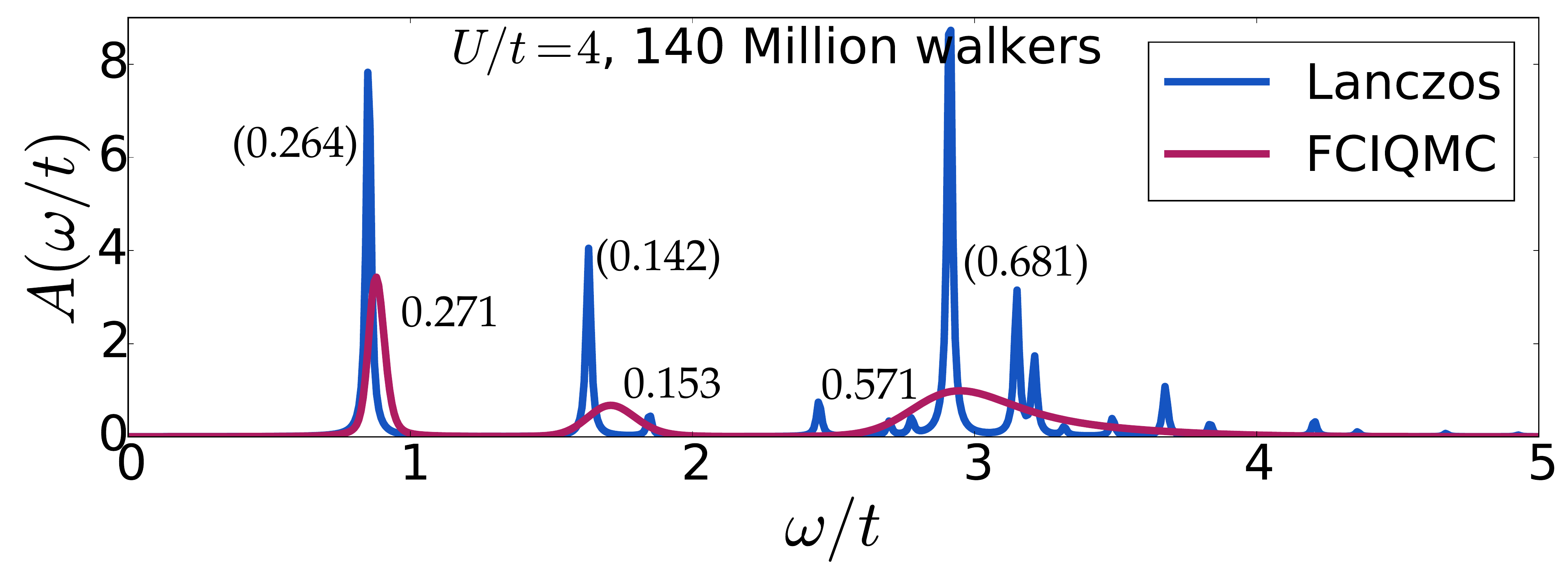}
\caption{Spectrum for the 18-site half-filled model at $U/t=4$ obtained
with $1.4\times 10^8$ walkers for $k=(0,0)$. Both the integrated weights of
the peaks of the FCIQMC spectrum as well as the corresponding integrated
weights of the Lanczos spectrum (bracketed) are displayed, showing reasonable agreement.}
\label{fig:HubU4}
\end{figure}

In addition to the study on the half-filled 18-site Hubbard model with
$U/t=2$, calculations on the same system with $U/t=4$ have been performed, of
which the resulting spectra are displayed in Fig.~\ref{fig:HubU4}.

\begin{figure}[t!]
\includegraphics[width=\columnwidth]{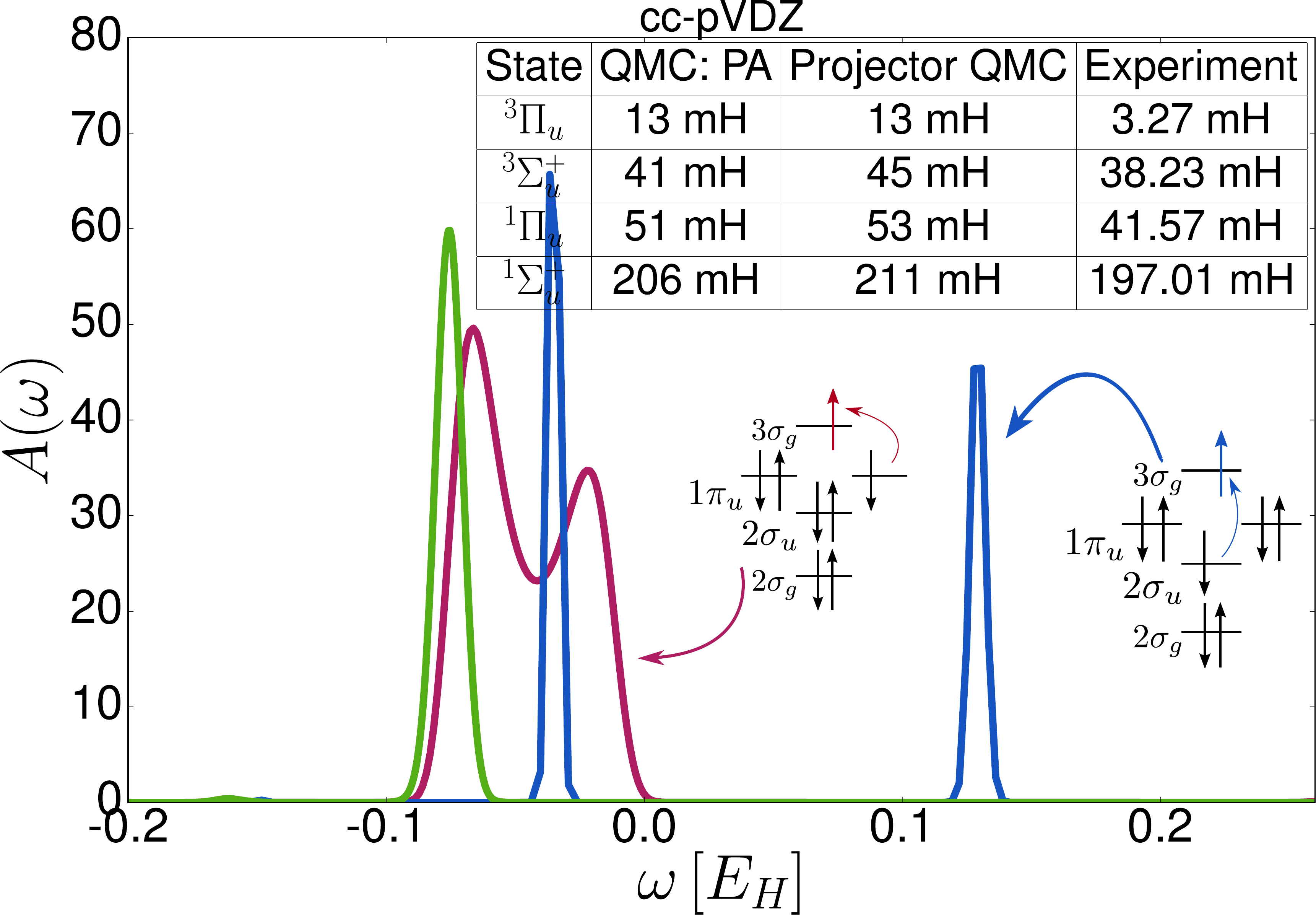}
\caption{Photo absorption spectra for the Carbon dimer in a cc-pVDZ basis set
  for a single excitation from the $2\sigma_u$(blue)/$1\pi_u$(red) to the
  $3\sigma_g$ orbital. Next to the real-time FCIQMC estimates of the
  excitation energies we also list the corresponding energies as obtained
  using excited-state i-FCIQMC method \cite{excited_state_neci} and the FCIQMC ground
state energy from \cite{BTCA2012}. The time step used is $\Delta t = 5 \times
10^{-3}$. Again, a rotation of time in the complex plane is performed, with an
angle of $\alpha \sim 0.33$, which is higher than for the larger basis sets
due to the larger time-step, leading to an increased broadening.}
\label{fig:C2DZ}
\end{figure}

For completeness, we also consider the Carbon dimer in a minimal cc-pVDZ basis
set consisting of 14 orbitals per atom in the frozen core approximation. The
Hilbert space size here is $\sim 10^8$, and photo absorption spectra can be
obtained analogously to the basis sets described in the main text, which are
shown in Fig.~\ref{fig:C2DZ}.

\end{document}